\documentclass[onecolumn,showpacs,preprint,amsmath,amssymb,aps,pra]{revtex4}

\usepackage{mathrsfs}
\usepackage{txfonts}
\usepackage{amsmath}
\usepackage{graphicx}
\usepackage{dcolumn}
\usepackage{bm}
\usepackage{ulem}

\begin{document}
\title[EP3 from anti-PT symmetry]{Higher-order exceptional points using lossfree negative-index materials}

\author{Xin-Zhe Zhang$^{1}$, Li-Ting Wu$^{2}$, Ru-Zhi Luo$^{1}$, and Jing Chen$^{1,3}$} \email{jchen4@nankai.edu.cn}

\address{$1$ MOE Key Laboratory of Weak-Light Nonlinear Photonics, School of Physics, Nankai University, Tianjin 300071, China \\
$2$ School of Information and Communication Engineering, Nanjing Institute of Technology, Nanjing 211167, China \\
$3$ Collaborative Innovation Center of Extreme Optics, Shanxi University, Taiyuan, Shanxi 030006, China
}

\date{\today}

\begin{abstract}
Negative-index materials (NIMs) support optical anti-parity-time (anti-$\mathcal{PT}$) symmetry even when they are lossless. Here we prove the feasibility in achieving higher-order exceptional points (EPs) in lossfree waveguide arrays by utilizing the anti-$\mathcal{PT}$ symmetry induced by NIM.  Numerical simulation about a third-order EP fits well with the coupled-mode theory. A scheme of achieving fourth-order EPs is also discussed. This work highlights the potential of lossfree NIMs in the study of non-Hermitian optics.

\end{abstract}

\pacs{123 }


\maketitle

\section{Introduction}

Exceptional points (EPs) refer to the singular degeneracies of non-Hermitian wave/quantum systems \cite{R01, R02, R03, R04, R05, R06, R07}, where all the eigenvalues and the eigenvectors of the effective Hamiltonian coalesce simultaneously. Its novel topology enables interesting mode switching behaviors when circling around it \cite{R06, R07, R08, R09, R10}. Furthermore, the coalescent eigenfunction at EPs is very sensitive to tiny perturbation in the effective Hamiltonian, so high-sensitivity applications \cite{R11, R12, R13, R14} can be envisioned. The high sensitivity is also associated with stopped-light effect and enhanced density of states, and can be utilized for other interesting purposes such as in realizing coherent absorbers and lasers \cite{R15, R16, R17}.

In most cases only the second-order EPs (EP2 for brief) are considered because EPs are generally believed to exist mainly in open systems. A good example is that in a parity-time ($\mathcal{PT}$) symmetric system that asks for delicate balance among spatially distributed gain and loss \cite{R01, R02}. If the order of EP is increased, it would provide more degrees of freedoms in artificially designing the topology, and the sensitivity can be further enhanced. Consequently, people have proposed many schemes in achieving higher-order EPs, e.g. in various kinds of photonic crystals, microcavities, and waveguides (WGs) \cite{R11, R12, R18, R19, R20, R21, R22, R23, R24, R25, R26, R27, R28, R29, R30}. However, the presence of gain and loss still dramatically hinders the transfer of EPs from a scientific curiosity to realistic applications of our daily life. Ways to access high-order EPs without resorting to gain and loss are thus desired. Such a target is, in principle, achievable because non-Hermitian physics covers many miscellaneous categories including but not limited to the $\mathcal{PT}$ symmetry. In addition, to achieve a non-Hermitian Hamiltonian, besides introducing imaginary components (gain and loss) to the diagonal elements, we can also just set the off-diagonal elements unequal. The later method does not require gain and loss, and the whole energy can be conserved. A good example is the recently demonstrated anti-$\mathcal{PT}$ symmetry associated with lossfree negative-index materials (NIMs) \cite{R31, R32, R33, R34, R35, R36}. Albeit the energy in the lossfree NIM-dielectric system is conserved \cite{R35, R36}, due to the backward propagation of field in NIM, the coupling between a NIM WG and an ordinary dielectric WG can be modeled by an anti-$\mathcal{PT}$ Hamiltonian of unequal off-diagonal elements. EP2s are shown to exist in the spectra of the guided waves \cite{R35, R36}.

In this article, we would prove the feasibility in realizing an optical third-order EP (EP3) by using the anti-$\mathcal{PT}$ symmetry induced by lossfree NIMs. An effective Hamiltonian is developed by using the coupled-mode theory in order to explain the existence of EP3. The transfer-matrix method (TMM) is utilized to numerically calculate eigensolutions of the guided mode, prove the existence of EP3, and reveal features of it. A scheme of realizing a fourth-order EP (EP4) is also discussed. This work proves that NIMs and the associated anti-$\mathcal{PT}$ symmetry have great potential in the study of non-Hermitian optics in lossfree environments for various applicable purposes.

\begin{figure}[htb]
\centerline{\includegraphics[width=10cm]{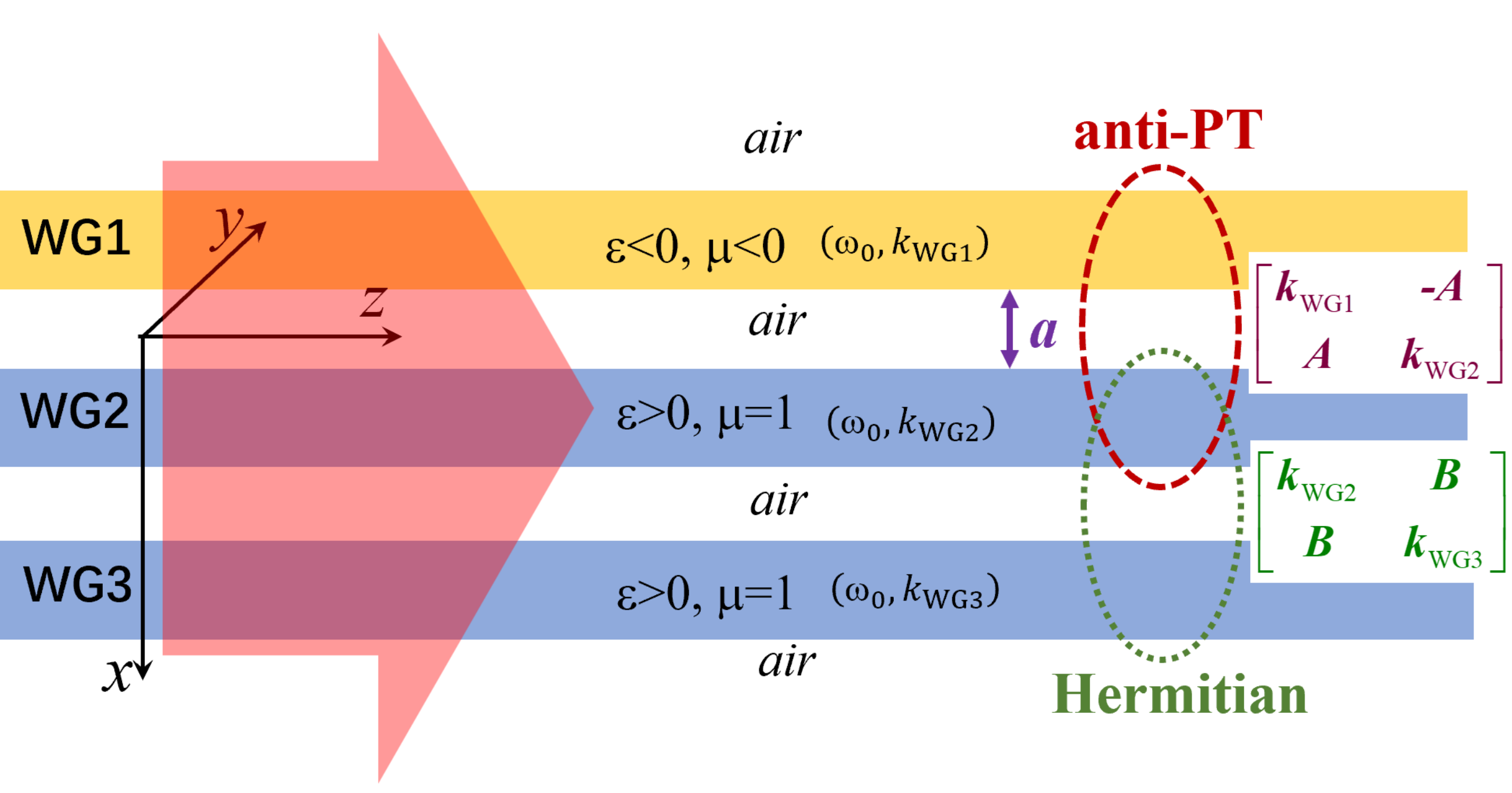}} \caption{ (Color
online) Schematic of the configuration under investigation, which contains three straight lossfree WGs. WG1 is made of a NIM with $\epsilon_{\rm NIM}=-3$ and $\mu_{\rm NIM}=-0.556$. WG2 and WG3 are made of dielectrics.}
\end{figure}

This article is organized as follows. In Section 2.1 we firstly propose the main concept of the coupled-WG structure and the physical mechanism of EP3 by using coupled-mode theory. In Section 2.2 we provide numerical calculation and analysis about the guided modes by using TMM.
We show that the Hamiltonian from the coupled-mode theory can explain main features of the results from TMM, and proves the observed degeneracy is indeed an EP3. The analysis also provides more detailed information about how the eigenmodes evolve near EP3. Discussion about the importance of this study is provided in Section 3. We also present a simple scheme of achieving an EP4 in this Section. Summary is made at the end of this article.

\section{Theory and Analysis}

\subsection{Structure and Effective Hamiltonian}

Let us consider the structure shown in Fig. 1. It contains three straight WGs surrounded by air. All the media in this structure are lossfree. The lower two WGs (WG2 and WG3) are made of dielectrics with $\epsilon>0$ and $\mu=1$. The top WG1 is made of a NIM with $\epsilon_{\rm NIM}<0$ and $\mu_{\rm NIM}<0$. Because NIM requires an intrinsic dispersion of $\partial (\epsilon_{\rm NIM}\omega) / \partial \omega>0$ and $\partial (\mu_{\rm NIM}\omega) / \partial \omega>0$ so as to give a positive energy density \cite{R31, R32, R33, R36}, in this article we would keep the angular frequency $\omega_0$ a constant, and test the variation of the wavevectors $\beta$ of the eigenmodes versus a geometric parameter of the structure.

As been discussed in \cite{R35, R36}, because the total energy should be conserved in this loss-free system, and the propagating directions of energy in the NIM and dielectric WGs are opposite to each other, the coupling between the top NIM WG1 and the adjacent dielectric WG2 is anti-$\mathcal{PT}$ symmetric \cite{R35, R36}. As for WG2 and WG3, their interaction can be described by using a Hermitian matrix. Only keeping the nearest-neighbor interaction, the coupled-mode theory gives an effective Hamiltonian on the eigensolutions $\beta$ of the guided mode

\begin{equation}
\left[
\begin{array}{ccc}
  k_{\rm WG1}  & -A & 0 \\
  A & k_{\rm WG2} & B \\
  0 & B& k_{\rm WG3} \\
  \end{array}\right]
\left[
\begin{array}{ccc}
   \psi_{1} \\
   \psi_{2} \\
   \psi_{3} \\
  \end{array}\right]=
\beta\left[
\begin{array}{ccc}
   \psi_{1} \\
   \psi_{2} \\
   \psi_{3} \\
  \end{array}\right],
\end{equation}
where $k_{\rm WG\it{i}}$ is the resonant wavevector of mode in separate WG$i$ ($i=1,2,3$), and $\psi_{i}$ represents an associated field component of it. Parameter $A$ represents the strength of the anti-$\mathcal{PT}$ coupling, and $B$ is the Hermitian coupling strength between WG2 and WG3. Both $A$ and $B$ are real.

Equation (1) generally has three solutions. Here let us assume the resonances in WG2 and WG3 are degenerated but different from that of WG1,
\begin{equation}
k_{\rm WG1}=k_0+3\delta, 
k_{\rm WG2}=k_{\rm WG3}=k_0,
\end{equation}
where the factor $3$ before $\delta$ is intentionally introduced in order to make below analysis concise. Substituting them into Eq. (1) and assume
\begin{equation}
y=\beta-k_0-\delta,
\end{equation}
the three solutions of $\beta$ can be found by solving
\begin{equation}
y^3+C_1y+C_0=0,
\end{equation}
where the coefficients are given by
\begin{equation}
C_1=A^2-B^2-3\delta^2,
C_0=(A^2+2B^2-2\delta^2)\delta.
\end{equation}
Because $A$, $B$, $k_0$ and $\delta$ are all real, $C_0$ and $C_1$ are also real valued.

Assuming the three solutions are $y_{1,2,3}$, Eq. (4) has many general properties such as $y_1+y_2+y_3=0$ and $y_1y_2y_3=-C_0$.
A notable feature is that it can support an EP3 with three identical solutions of $y=0$ when
\begin{equation}
C_1=C_0=0.
\end{equation}
Equation (6) can be satisfied only when the conditions of
\begin{equation}
\delta=0, A^2=B^2
\end{equation}
are met simultaneously. These are the existence conditions of EP3, the main conclusion of this article.

Because parameters $A$ and $B$ are tunable by managing the distances between adjacent WGs, to gain a deep insight about the formation of EP3 and the variation of the associated eigenvectors, let us check how the eigensolutions vary about EP3. Substituting $k_{WG1}=k_{WG2}=k_{WG3}=k_0$
into Eq. (1) we can get
\begin{equation}
(\beta-k_0)^3+(\beta-k_0)(A^2-B^2)=0.
\end{equation}
Assuming all the solutions are real, we can sort them in ascending order. The solution $\beta_{2}$ is between the other two and is given by
\begin{equation}
\beta_{2}=k_0,
\end{equation}
which is a constant and does not depend on the values of $A$ or $B$. The other two solutions are given by
\begin{equation}
\beta_{1,3}=k_0\mp\sqrt{B^2-A^2},
\end{equation}
which are complex (real) in the region of broken (exact) phase when $B^2<A^2$ ($B^2>A^2$). As for the eigenvectors, from Eq. (1) we can find
\begin{equation}
\Psi_{2}=\frac{1}{\sqrt{A^2+B^2}}\left[B, 0, -A\right]^T, 
\Psi_{1,3}=\frac{1}{\sqrt{2}B}\left[A, \pm\sqrt{B^2-A^2}, -B\right]^T,
\end{equation}
respectively.

Once the condition of $A^2=B^2$ is satisfied, a degeneracy takes place, where all the three eigensolutions coalesce together. This degeneracy is an EP3, and the eigensolution and eigenvector are given by
\begin{equation}
\beta_{\rm EP3}=k_0,
\Psi_{\rm EP3}=\frac{1}{\sqrt{2}}\left[1, 0, -{\rm sign}\left(\frac{B}{A}\right)\right]^T,
\end{equation}
respectively, where the function sign($x$) returns 1 ($-1$) when  $x>0$ $(x<0)$.

If the two conditions of Eq. (6) are not satisfied simultanously, e. g.  when $\delta\neq0$, we could not access EP3. Now the system at most supports an EP2. One solution of $\beta$ is always real, and the other two solutions are determined by
\begin{equation}
\Delta=\left(\frac{C_0}{2}\right)^2+\left(\frac{C_1}{3}\right)^3.
\end{equation}
When $\Delta=0$, the two solutions are identical and an EP2 is achieved. When moving from $\Delta=0$, for example, by changing the distance between two adjacent WGs so as to modify $A$ or $B$, the system would enter either the exact or the broken non-Hermitian phase. If $\Delta>0$, two complex conjugate solutions are achieved, and the system is within the broken phase. If $\Delta<0$, two real solutions with different values are found, and the system is within the exact phase. In the next section we would demonstrate this phenomenon.

\begin{figure}[htb]
\centerline{\includegraphics[width=14cm]{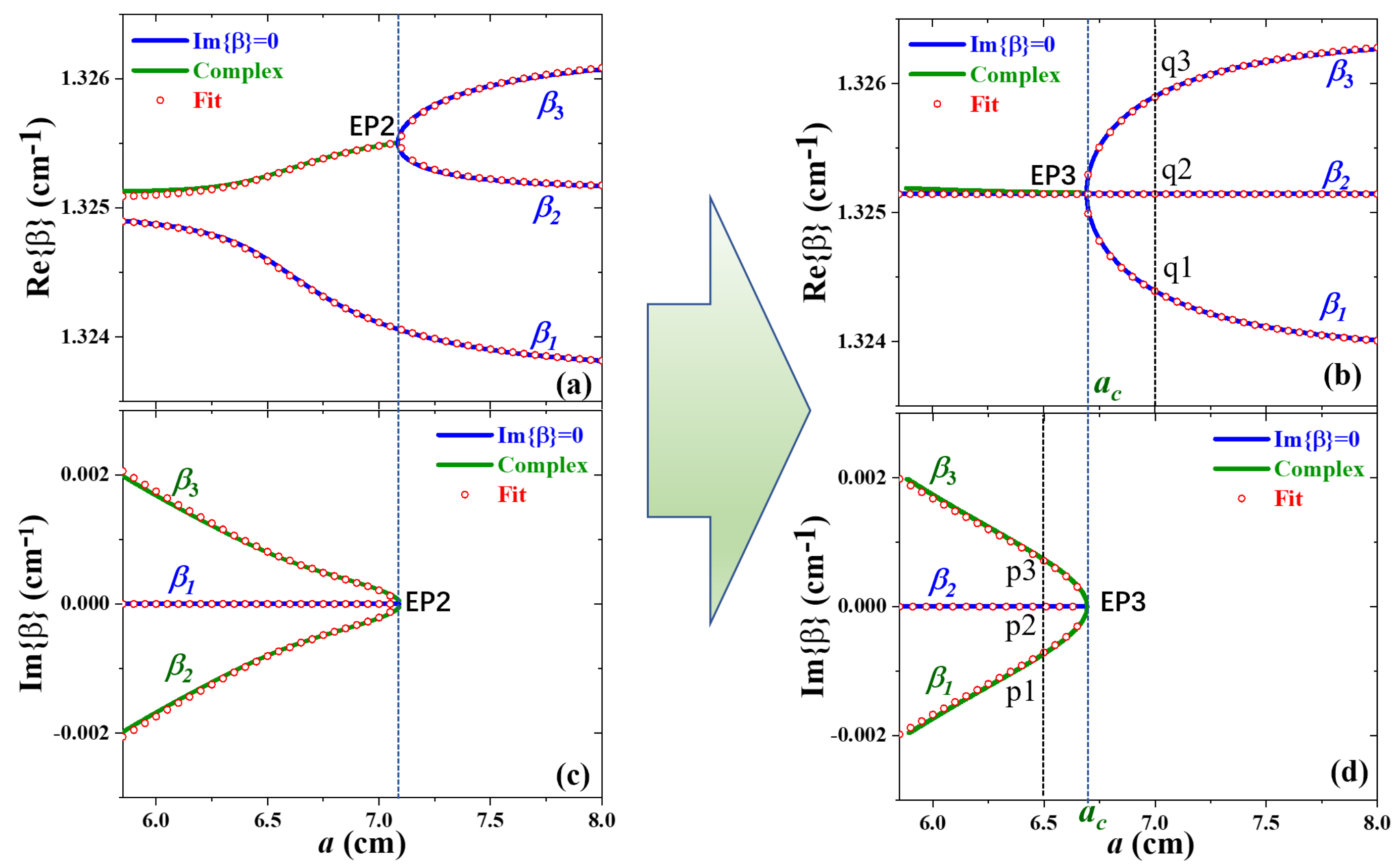}}
\caption{ (Color online) Variation of the eigensolutions $\beta$ versus the distance $a$ by using TMM. (a,c) In general we can always get an EP2, where the other dispersion curve is real and always persists. (b, d) By finely tuning parameters of the structure we can get an EP3. Red circles are the best fitted results by using Eq. (1).}
\end{figure}

\subsection{Numerical Simulation and Analysis}

Above theory about the eigensolutions and eigenvectors is based on the effective Hamiltonian from the coupled-mode theory. When studying the guided mode in coupled WGs we should still resort to methods based on Maxwell's equations. Here we analyze the transverse-electrical modes in the structure by using TMM. In each layer the field is expressed as $E_y=[E_+\exp(jk_ix) +E_-\exp(-jk_ix)]\exp(-j\beta z) $, where $k_i^2+\beta^2=\epsilon_i\mu_i\omega_0^2/c^2$, $c$ is the speed of light. The transmission/reflection properties of the structure are summarized by $[E_t, 0]^T=\bm{M}[E_0, E_r]^T$, where $E_0$, $E_t$ and $E_r$ are the incident, transmitted, and reflected fields, $\bm{M}$ is the transfer matrix. At the top and bottom air media surrounding the WGs, field is required to exponentially decay away ($\beta>\omega_0/c$) so that out-going boundary conditions are met. Wave-guiding is satisfied by the self-sustained condition of $M_{22}=0$ \cite{R37}, so that $E_r$ and $E_t$ are nonzero even when no incidence is present ($E_0=0$). In the phase-broken regions where $\beta$ is complex, similar process can be utilized as well by scanning $\beta$ in the two dimensional space spanned by ${\rm Re}\{\beta\}$ and ${\rm Im}\{\beta\}$. As for the associated distribution of fields, they can be calculated also by TMM when an eigensolution is found.

The geometric and optical parameters of the structure are set as follows. The thicknesses of all WGs are 4 cm. The distance between WG2 and WG3 is set to be 6 cm so that $B$ is a constant here. The distance $a$ between WG1 and WG2, which determines the value of $A$, is variable in our study. NIM is assumed to be the documented one with $\epsilon_{\rm NIM}=1-\omega_e^2/\omega^2$ and $\mu_{\rm NIM}=1-F\omega^2/(\omega^2-\omega_m^2)$, where $\omega_e=2\pi\times10$ GHz, $\omega_m=2\pi\times4$ GHz, and $F=0.56$ \cite{R32, R33}. At $\omega_0=2\pi\times5$ GHz  (a free-space wavelength of 6 cm) the dispersion gives $\epsilon_{\rm NIM}=-3$ and $\mu_{\rm NIM}=-0.556$.

Figs. 2(a) and 2(c) display the spectra of $\beta$ verses $a$ when WG2 and WG3 are made of the same dielectric of $\epsilon=4.504$. In this case $k_{\rm WG2}=k_{\rm WG3}\neq k_{\rm WG1}$ so we could only get an EP2. From the curves we can see the lower branch $\beta_1$ is always real, but the upper two branches of $\beta_2$ and $\beta_3$ coalesce together in forming an EP2 when $a=7.09$ cm. When $a$ is smaller than 7.09 cm, the imaginary parts of these two branches are no longer zero, and the non-Hermitian phase is broken.

In order to realize EP3, we should finely tune the parameters of the structure. We find that when $\epsilon$ of the two dielectric WGs are modified to 4.50465083, an EP3 is achieved, see Figs. 2(b) and 2(d). From Fig. 2(b) we can see just this tiny $0.015\%$ tuning would sharply modify the dispersion curves. Now the initial lower branch $\beta_1$ coalesce with the upper two branches $\beta_{2,3}$ at a single point at $a_c=6.69$ cm and forms an EP3. When $a$ becomes smaller, one branch of $\beta$ is always real, while the other two become complex. The middle branch is always real and is almost a constant against $a$. Note that due to the coalescence at EP3, it is impossible to find the exact one-to-one correspondence among the three branches of $\beta$ in the regions of broken and exact phases, so the complex branches are still labeled in ascending order of $\rm Im\{\beta\}$.

Above numerical calculations are based on TMM, which is rigorous. Before discussing characteristics of EP3 we should first check how they fit with the coupled-mode theory of Eq. (1). To do it, we perform a best fitting about the dispersion curves by numerical solving Eq. (1). We first fit the curves shown in Figs. 2(b) and 2(d) because they are govern by $\delta=0$ and require less fitting parameters. The value of $k_0$ is given by the numerical value of $\beta_2=1.325144$ cm$^{-1}$ obtained from TMM. As for A and B, following \cite{R36} we assume
\begin{equation}
A=B\exp\left(-\frac{a-a_c}{L_d}\right),
\end{equation}
where $a_c=6.69$ cm is the position of EP3, and $L_d$ is the decay length. The best fitting then requires $B=1.22\times 10^{-3}$ cm$^{-1}$ and $L_d=1.3$ cm. From the red circles of Figs. 2(b) and 2(d) we can see the coupled-mode theory predicts most features of the dispersion curves from TMM. The only minor discrepancy is about $\rm Re\{\beta\}$ in the broken phase region, which might be fixed by considering high-order dispersion effect.

We then try to fit the results shown in Figs. 2(a) and 2(c). Since the only difference of them from Figs. 2(b) and 2(d) is that the resonances in WG2 and WG3 are perturbed by the same amount, we would maintain all the parameters about Figs. 2(b) and 2(d) unchanged, and only introduce a tiny perturbation to $k_{\rm WG2, WG3}$, that $k_{\rm WG2}=k_{\rm WG3}=k_0+\Delta k$. A best fitting is achieved when $\Delta k=-1.8\times 10^{-4}$ cm$^{-1}$, as shown as the red circles. We can see once again they fit well with these from TMM. It confirms that the coupled-mode theory of Eq. (1) is an accurate model about the guided modes in the coupled WGs.

\begin{figure}[htb]
\centerline{\includegraphics[width=12cm]{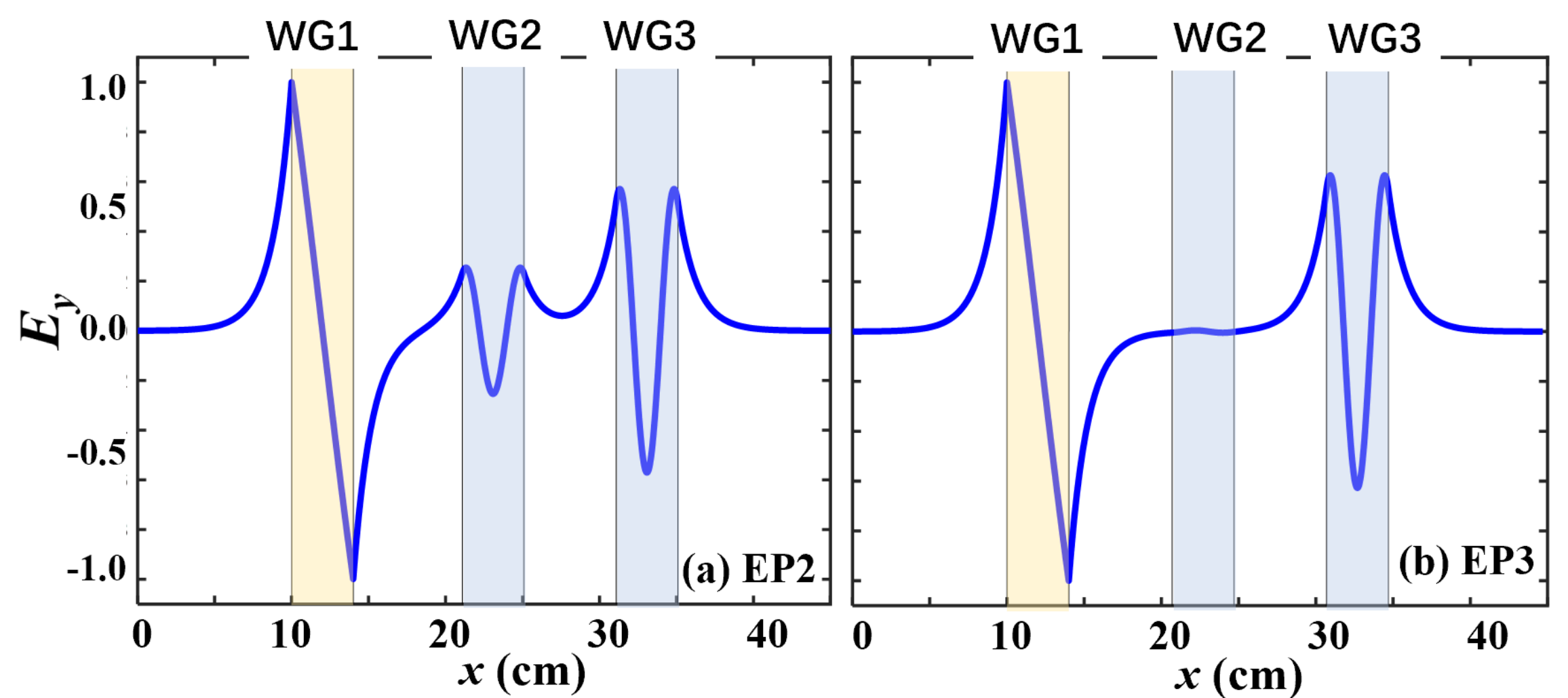}}
\caption{ (Color online) Distributions of field $E_y$ at (a) EP2 and (b) EP3 shown in Fig. 2.}
\end{figure}

Now we can try to analyze the formation of EP3 and the associated characteristics of it. We first calculate the distributions of $E_y$ at EP2 and EP3 of Fig. 2, and show them in Fig. 3. We can see EP3 is sharply difference from EP2 because the field inside WG2 is nearly zero, which is in agreement with Eq. (12). The patterns of fields in WG1 and WG3 are similar in these two scenarios.

\begin{figure}[htb]
\centerline{\includegraphics[width=15cm]{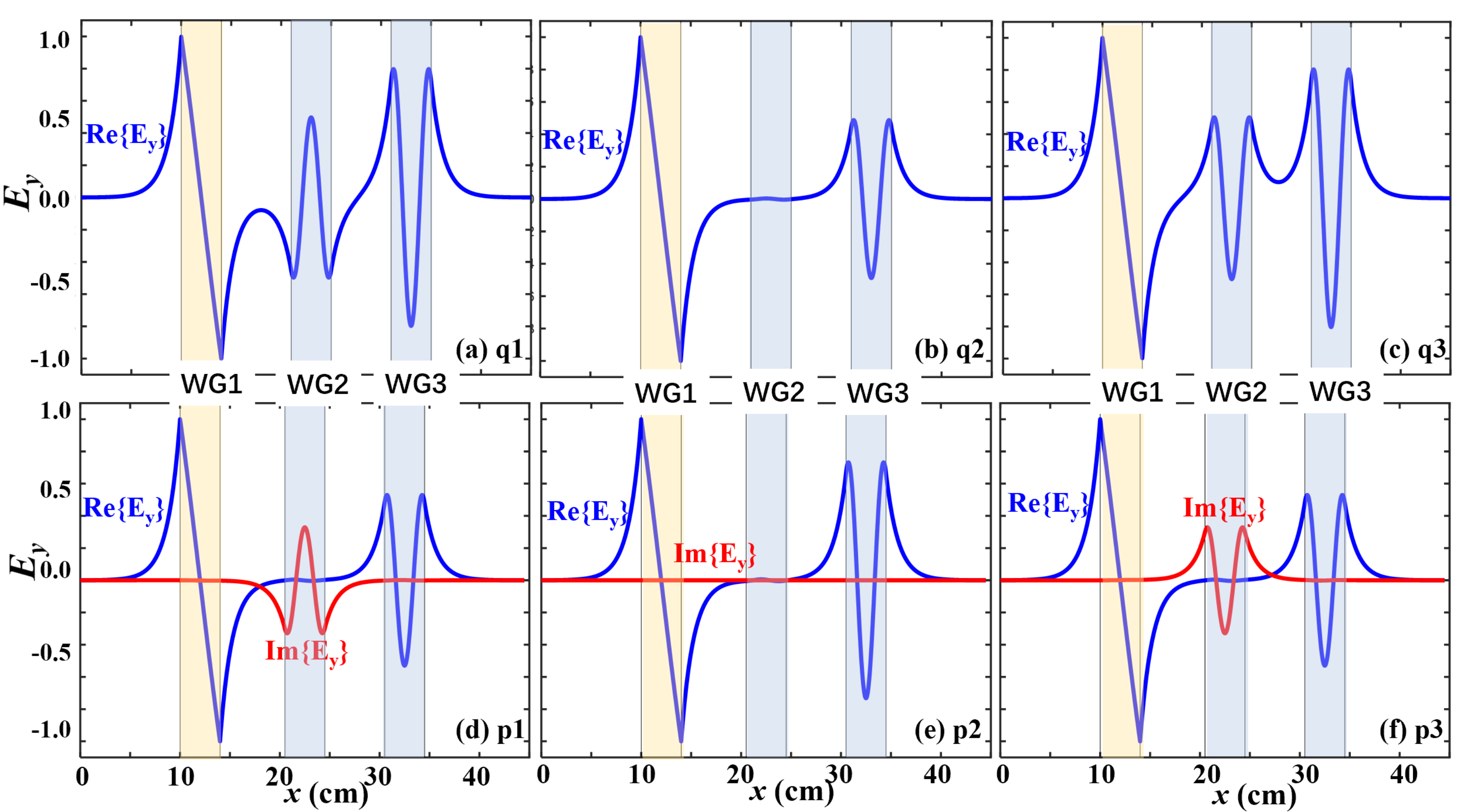}}
\caption{ (Color online)  Distributions of field $E_y$ at the chosen points of Figs. 2(b) and Fig. 2(d) supporting EP3.}
\end{figure}

We then pay attention to the structure supporting EP3, and calculate the distribution of $E_y$ at six points around EP3 in the dispersion curves of Fig. 2(b) and 2(d). The result are shown in Fig. 4. For the three points beyond EP3 ($a>a_c$), all the solutions of $\beta$ are real. As a consequence, the field $E_y$ is real, see Figs. 4(a) to 4(c). A notable feature is that for the point $p2$ in the middle branch $\beta_2$, the field inside WG2 is very weak, similar to EP3. This phenomenon is also in agreement with Eq. (11) .

When $a<a_c$, two branches of $\beta$ becomes complex. Now the field $E_y$ becomes complex, so in Figs. 4(d) to 4(f) we also plots the imaginary parts of it. From the distribution of fields we can see in these broken phase scenario, the patterns of the real part of $E_y$ are almost identical with each other (also with that of EP3). The differences in the distribution of $E_y$ are carried by the imaginary part. For the two complex-$\beta$ points, the imaginary parts of $E_y$ are opposite with each other and is especially strong around WG2. This phenomenon can be explained by using Eq. (11), that in the region of broken phase, the complex amplitude of the basic vector $\psi_{2}$ in $\Psi_{1,3}$ is purely complex and possesses a $\pi/2$ or $3\pi/2$ phase difference from these of the basic vectors $\psi_{1,3}$.

To prove that at EP3 the three dispersion branches $\beta$ coalesce together, we test whether their eigenvectors follow the prediction of Eq. (11). It can be done by checking how the relative magnitude of field inside WG2 varies with $a$. Here we assume that $\psi_i$ in the eigenvector represents $E_y$ inside WG$i$ ($i=1,2,3$), calculate the integral of $|E|^2$ inside each WG, and find the ratio $\alpha$ of $|E|^2$ confined in WG2 by using
\begin{equation}
\alpha=\frac{\int_{\rm WG2}|E_y|^2dx}{\int_{\rm WG1}|E_y|^2dx+\int_{\rm WG2}|E_y|^2dx+\int_{\rm WG3}|E_y|^2dx}.
\end{equation}
Albeit this approach is very rough, e.g. because the fields outside WGs are ignored, the variation of $\alpha$ versus $a$ shown in Fig. 5(a) agrees well with the dependence of $|\psi_2|^2$ on $A$ given by Eq. (11). For modes in the middle branch $\beta_2$, $\alpha$ is nearly zero, which is in agreement with the null $\psi_2$ given by Eq. (12). As for the other two branches, when $a$ is greater than $a_c$, the magnitude of $A$ is much smaller than $B$, i.e. $|A|\ll|B|$. The eigenvectors are close to $[0, \pm1,1]^T$, and $\alpha$ approaches $50\%$. When $a$ decreases so that $A$ becomes comparable with $B$,  $\alpha$ decreases sharply to zero. At $a_c$ where $A=B$, as expected, a coalescence takes place. Because the curves of $\alpha$ are continuously varying, we can make the conclusion that this point at $a=a_c$ is not an accidental degeneracy of EP2 with other modes but a standard EP3.

\begin{figure}[htb]
\centerline{\includegraphics[width=14cm]{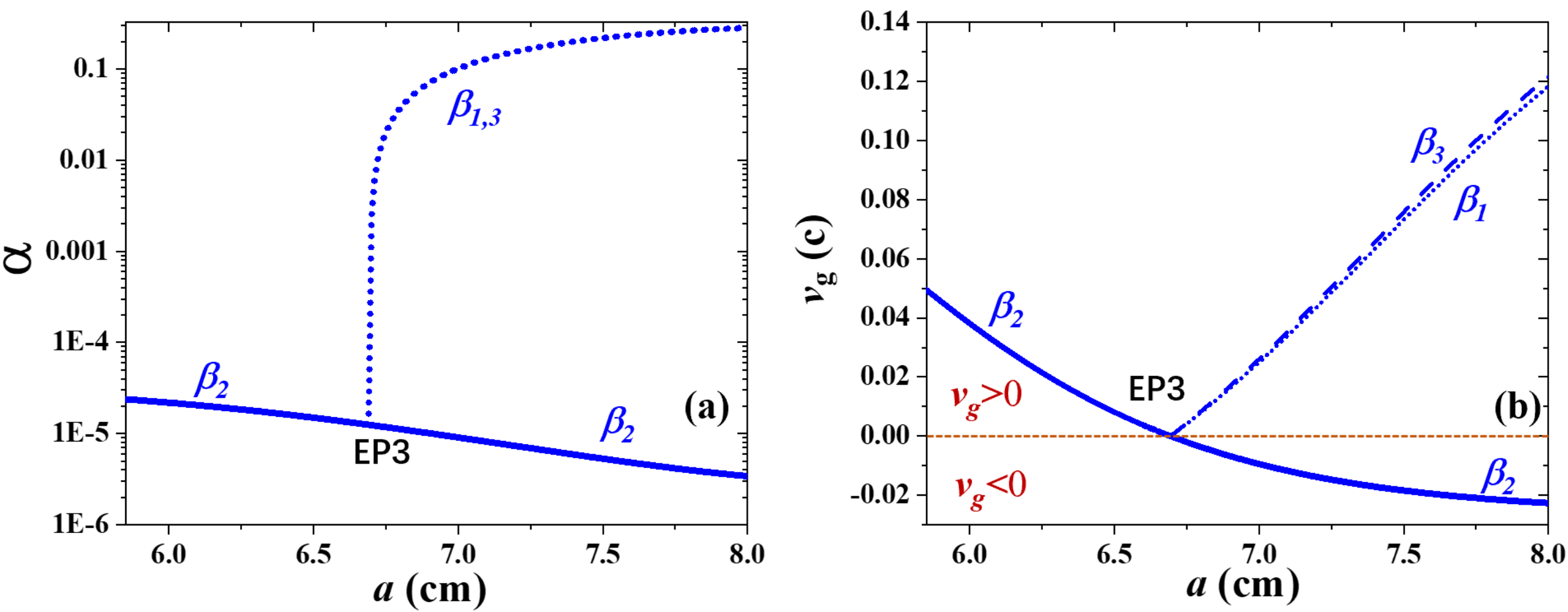}}
\caption{ (Color online)  (a) Ratio of field localized inside WG2, where the $y$ axis is plotted logarithmic. (b) Group velocities $v_g$ of the three branches. Results about complex $\beta$ are not shown because the associated $\alpha$ and $v_g$ are not well defined.}
\end{figure}

Since EPs in $\mathcal{PT}$-symmetric WGs stop light \cite{R15, R16}, we also calculate the group velocity $v_g$ of the three branches. Here $v_g$ is given by the ratio of the Poynting vector $S_z=-\int E_yH_x^*dx/2$ to the energy density $W=\int wdx$ via the formula of $v_g=S_z/W$. When calculating the energy density we have adapted $w=\epsilon_0\partial (\epsilon_{\rm NIM}\omega) / \partial \omega |E|^2/4+\mu_0\partial (\mu_{\rm NIM}\omega) / \partial \omega |H|^2/4$ in order to guarantee a positive energy. At 5 GHz, the utilized dispersion of NIM gives $\partial (\epsilon_{\rm NIM}\omega) / \partial \omega=5$ and $\partial (\mu_{\rm NIM}\omega) / \partial \omega=4.975$.

Figure 5(b) shows the variation of $v_g$ versus $a$ of the three branches $\beta$. When $a>a_c$ so that $A<B$, the $\beta_2$ branch possesses a negative group velocity. The reason is that for modes in this branch, over half of the field is localized in WG1 that is NIM and supports backward propagation of field. As for the other two branches, they are generally positive. When approaching $a_c$, all the branches coalesce together to $v_g=0$, which implies that at $a_c$ the backward energy flux in the NIM WG balances the forward one in the dielectric WGs and the surrounding air. This phenomenon is also demonstrated in \cite{R36} about EPs in NIM-dielectric WGs. Once again, Fig. 5 confirms that at $a_c$ an EP3 is achieved.

\section{Discussion}

Up to now we have proved that it is feasible to achieve an EP3 in a lossfree WG system containing NIM. This kind of EP3 is produced via the hybridization of the anti-$\mathcal{PT}$ symmetric coupling in the NIM-dielectric pair and the Hermitian coupling in the dielectric-dielectric pair. No loss or gain is required. This work highlights the great potential of NIM in overcoming the obstacles of ordinary non-Hermitian optics, and the possibilities of combining anti-$\mathcal{PT}$, $\mathcal{PT}$, and Hermitian couplings for various purposes. Albeit in this article we only consider NIM-dielectric-dielectric configuration, EP3 can be achieved in the dielectric-NIM-NIM configuration as well because the coupling in the NIM-NIM pair is also Hermitian.

Our work provides a useful route in designing lossfree systems in achieving EPs with further increased orders. For example, here we can propose a schematic structural design for an EP4 by using two dielectric and two NIM WGs. Arranging these four WGs parallel in the dielectric-dielectric-NIM-NIM order, and assuming their resonant wavevectors at $\omega_0$ are $k_1$, $k_2$, $k_1$, and $k_2$, respectively, the effective Hamiltonian based on the coupled-mode theory with nearest-neighbor interaction can be expressed as
\begin{equation}
\left[
\begin{array}{cccc}
  k_1  & A & 0 & 0 \\
  A & k_2 & B &0 \\
  0 & -B& k_1& C \\
  0 & 0& C& k_2 \\
  \end{array}\right]
\left[
\begin{array}{cccc}
   \psi_1 \\
   \psi_2 \\
   \psi_3 \\
   \psi_4 \\
  \end{array}\right]=\beta\left[
\begin{array}{cccc}
   \psi_1 \\
   \psi_2 \\
   \psi_3 \\
   \psi_4 \\
  \end{array}\right],
\end{equation}
where $A$, $B$, and $C$ are all real for simplification. Assuming
\begin{equation}
k_{1,2}=k_0\pm\delta,
\end{equation}
it is then easy to prove that under the conditions of
\begin{equation}
B=\pm(|A|+|C|),
\delta^2=|AC|,
\end{equation}
all the four eigensolutions coalesce together in forming an EP4 at
\begin{equation}
\beta_{\rm EP4}=k_0.
\end{equation}
The hierarchical construction of higher-order EPs by using above proposed method deserves a further discussion.

\section{Conclusion}

In summary, here we show it is feasible to achieve an EP3 in loss-free WGs by utilizing the anti-$\mathcal{PT}$ symmetry induced by NIM. TMM simulation agrees well with the prediction of coupled-mode theory. A structure in supporting EP4 is also designed. This work highlights the great potential of loss-free NIM, together with the associated anti-$\mathcal{PT}$ symmetry, in the study of non-Hermitian optics.

This work was supported by the Natural National Science Foundation of China (NSFC) (12104227, 12274241), and the Scientific Research Foundation of Nanjing Institute of Technology (YKJ202021).

\end{document}